\documentclass[12pt]{article}
\usepackage{epsfig}
\begin{document}
\noindent
{\Large \bf How to complete light meson spectroscopy  to
$M = 2410$ MeV/c$^2$}
\vskip 5mm
\noindent
{D V Bugg \footnote {email: d.bugg@rl.ac.uk}} \\
{Physics Department, Queen Mary, University of
London, London E1\,4NS, UK} \\
\vskip 1cm

\begin {abstract}
A measurement of transverse polarisation in $\bar pp \to $ all-neutral
final states would almost certainly determine a complete set of
partial wave amplitudes over the mass range 1910 to 2410 MeV.
This should identify all resonances in this mass range.
The experiment is technically straightforward and cheap by present
standards.
\end {abstract}
\vskip 1cm

\noindent
One simple experiment has an excellent chance of completing
spectroscopy of light mesons up to a mass of 2410 MeV.
It requires a measurement of transverse polarisation in reactions
\begin {eqnarray}
\bar pp &\to& \pi ^0\eta, \, 3\pi^0 \,\, {\rm and} \,\,\, \pi ^0 \eta \eta,
\qquad \eta \to \gamma \gamma \\
&\to& \omega \pi ^0 \,\,\, {\rm and}\,\,\, \omega \pi ^0\eta, \qquad
\omega \to \pi^0 \gamma \\
&\to& \omega \eta \,\,\, {\rm and} \,\,\, \omega \pi^0 \pi^0, \qquad
\omega \to \pi^0 \gamma
\end {eqnarray}
This is a formation experiment of the form $\bar pp \to
{\rm resonance} \to A + B$, where $A$ and $B$ are decay channels.

Data for differential cross sections already exist from the
Crystal Barrel experiment at LEAR at eight beam momentum from 900 to
1940 MeV/c.
Results and technical details are reviewed in Ref. \cite {review}.
There are also extensive measurements of polarisation in
$\bar pp \to \pi ^+\pi ^-$ from two earlier CERN experiments
\cite {Eisenhandler} \cite {Hasan}.
The importance of the polarisation data are illustrated vividly by
the fact that a unique set of partial wave amplitudes is found for
$I=0$ and $C = +1$.
Simulations from existing data show that reaction (1) would
likewise give a unique set of amplitudes for $I=1$ $C = +1$; reactions
(2) and (3) would do the same for both $C = -1$ states.
There would also be high statistics for $\bar pp \to
\pi ^0\pi ^0$, $\eta \eta$ and $\eta \pi ^0\pi ^0$; polarisation
data for these channels would check the present partial wave analysis 
for $I=0$, $C=+1$.

The $\bar pp$ system contains singlet $S$ and triplet $T$ spin
configurations; $d\sigma /d\Omega = |S|^2 + |T|^2$.
The formula for $P_Y d\sigma /d\Omega$ is $Tr(A^*\sigma _Y A)$,
where $\sigma$ is the Pauli matrix and $A$ the amplitude.
For polarisation normal to the production plane, $Pd\sigma /d\Omega$
determines the imaginary part of interferences; for sideways spin
(in the plane of scattering), it determines the real part of exactly
the same interferences.
Sideways polarisation $P_S$ is zero for 2-body final states because
momentum conservation demands that initial and final states lie in a
plane.
But for 3-body final states it is non-zero; it depends on
$\sin \phi$, where $\phi$ is the azimuthal angle between the plane of
polarisation and the decay plane of the 3-body final state.

A major virtue of polarisation data is that they are phase sensitive. 
Without polarisation data, there are always twofold ambiguities
in relative phases between amplitudes.
The constraint of analyticity removes some of these ambiguities, but
not all.
With polarisation, they are eliminated; Argand
diagrams for amplitudes can be traced unambiguously as a function of
beam momentum.
The phase sensitivity of polarisation data reduces errors on
measured masses and widths by typically a factor 2.
A further important point is that there are interferences between 
singlet and triplet states.
It is already clear there are large amplitudes from $f_4(2050)$,
$f_4(2300)$, $\rho_3(1982)$, $a_4(2005)$, $a_4(2255)$,
$\rho_3(1982)$ and $\rho_3(2260)$.
These serve as powerful interferometers for the
determination of small partial waves.
With the aid of dispersion relations or analytic functions fitting
the data, the partial wave amplitudes are unique without further
measurement of $A$ or $R$ parameters, as in $\pi N$ scattering.

A further crucial point is that triplet states such as $^3P_2$ and 
$^3F_2$ have orthogonal combinations of Clebsch-Gordan coefficients;
they are cleanly separated by polarisation data and 
$d\sigma /d\Omega$.
This separation is presently vital missing information for all of 
reactions (1)-(3).

A  frozen-spin target is required with the cryostat along the beam
direction.
Existing technology is perfectly adequate.
A Monte Carlo simulation based on existing data shows that
backgrounds from heavy nuclei are $\sim 10\%$; experience in
Ref. \cite {Hasan} and experiments at LAMPF \cite {Shypit}
confirm this.
A modest beam intensity of $\sim 6 \times 10^4$ $\bar p
/$s and a 3cm target of NH$_3$ gives $\sim  100$ good events/s.
Statistics of $\sim 50$K events per channel are required; this is 
hardest to achieve in the channel $\omega \eta$, but is possible with 
a running time of $\sim 10 $ days/momentum (assuming $70\%$ running
efficiency). Twelve beam momenta are required.

Angular coverage of 98\% of $4\pi$ is vital, because of
partial waves up to $L=5$ in the initial $\bar pp$ state.
The existing Crystal Barrel detector needs to
complete its present series of experiments on photoproduction
first, but would then be ideal for the required measurements
on meson spectroscopy.

Analysis of the mass range 1910 to 2050 MeV requires further
measurements of $d\sigma/d\Omega$ at $\sim 360$ MeV/c (the lowest 
beam momentum without stopping $\bar p$ in the target), $\sim 470$, 
600 and 750 MeV/c; these data were scheduled to be taken at LEAR,
but the machine closed just 2 weeks before the data could be  
taken.
Measurements from a polarised gas-jet target are unrealistic since
$d\sigma /d\Omega$ must be normalised accurately between momenta; 
also the geometry of the detector would pose serious problems.
Therefore an extracted beam such as those which existed at LEAR is
necessary, but is  technically not difficult.
Modifications to the existing Crystal Barrel detector and polarised
target would be minor, so this is not an expensive experiment.

A detail which is clear from existing Crystal Barrel data is
that $\bar pp$ interactions produce $s\bar s$ states only very
weakly.
Data on $\bar pp \to \pi ^0 \pi ^0$, $\eta \eta$ and $\eta \eta '$
determine mixing angles of states observed in these channels
between $n\bar n$ and $s\bar s$ \cite {Mixing};
they are mostly zero within experimental errors.
Even the prominent $f_2(1525)$ is hard to detect in $\bar pp$
interactions.

It is already known that high mass states have strong decays 
to final states such as $\pi \omega (1650)$ and 
$\pi \omega _3(1670)$.
There is therefore a hope that similar decays would reveal the
states presently missing in the mass range 1600-1900 MeV, but no
guarantee. 
There are missing $^3D_2$ and $^1S_0$ states for both isospins
and also the exotic $I=0$ $J^{PC}=1^{-+}$ state.
A general remark is that production experiments of the form
$\pi N \to X + N$ have the serious disadvantage that the exchanged
meson can have non-zero spin, necessitating the determination of 
both the exchanged spin and the spin of X.
Generally this is impossible or at best guesswork, hence explaining 
why experiments of this type have not been able to observe most of 
the states above 1900 MeV.

\begin{figure}[htb]
\begin{center}
\vskip -8mm
\epsfig{file=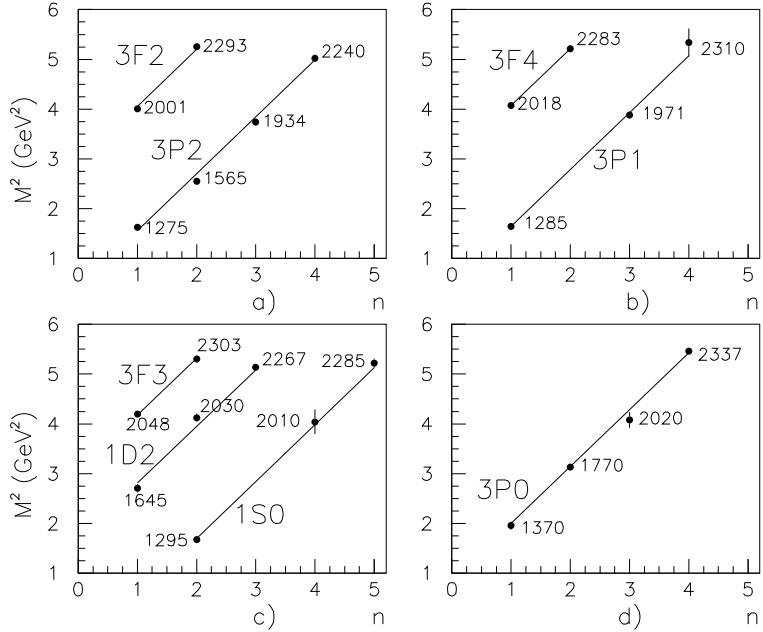,width=12.5cm}
\vskip -9mm
\caption {Trajectories of $I=0$, $C = +1$ mesons; $n$ is the radial
excitation number; masses are shown in MeV.}
\end{center}
\end{figure}

Confinement is one of the fundamental phase transitions of physics.
Completing the spectroscopy of light mesons and baryons should
therefore be a fundamental priority.
It is already clear that Chiral Symmetry Breaking plays a decisive
role at low masses.
At high masses, there is a striking regularity of observed states
illustrated in Fig. 1 for $I=0$ $C = +1$ states.
All of these states except the $^3P_0$ state (and one $^1G_4$ state,
not shown) have been observed in at least three channels of data, 
see Ref. \cite {review} for details.
They are listed by the Particle Data Group under `Other Light 
Mesons' \cite {PDG} with the remark that they have been observed
only by a single group and thus need confirmation.
Confirmation is indeed needed, and could be achieved in the 
proposed experiment. 
Several of the states in the regular listings have also been 
identified only by one group and likewise need confirmation.

Existing states fall close to parity doubling, e.g. $J^{PC}=2^{-+}$
and $2^{++}$, so an understanding of the extent of Chiral Symmetry 
restoration at high masses is imporant.
Glozman \cite {Glozman} argues in favour of full restoration,
on the basis that quarks are highly relativistic at such momenta,
so $J$ should be a good quantum number.
However, present data show definite mass shifts of $\sim 80$ MeV
between F and P states and $\sim 40$ MeV between  F and D states.
This leads Afonin \cite {Afonin} to argue that orbital angular
momentum $L$ is a better label, as in the hydrogen atom.
The physical explanation may be that $L$ is carried by the rotating
flux-tube, which breaks above a certain value of $L$. 
More precise determinations of masses and widths would clarify the
experimental situation greatly.

In summary, this is a straightforward experiment and the technology
exists. It should be done.

\end  {document}